\documentclass[12pt]{article}%
\usepackage{amsmath}
\usepackage{amsfonts}
\usepackage{amssymb}
\usepackage{graphicx}%
\providecommand{\U}[1]{\protect\rule{.1in}{.1in}}
\begin{document}
\begin{titlepage}
%\begin{flushright}
%2008/3/2
%\end{flushright}
\ \\
\begin{center}
\LARGE
{\bf
Ground-State Entanglement\\
gives birth to\\
Quantum Energy Teleportation
}
\end{center}
\ \\
\begin{center}
Contribution to Proceedings of International Conference on Quantum Communication and Quantum Networking, Oct. 26-30, Italy.
\end{center}
\begin{center}
\large{
Masahiro Hotta
}\\
\ \\
\ \\
{\it
Department of Physics, Faculty of Science, Tohoku University,\\
Sendai, 980-8578, Japan\\
hotta@tuhep.phys.tohoku.ac.jp
}
\end{center}
\begin{abstract}
Ground-state entanglement induces emergence of negative-energy-density regions in quantum systems
by squeezing zero-point oscillation, keeping total energy of the systems nonnegative.
By use of the negativity of quantum energy density, protocols of quantum energy teleportation are
proposed that transport energy to distant sites by local operations and classical communication.
The energy is teleported without breaking any physical laws including causality and local energy conservation.
Because intermediate subsystems of the energy transfer channel are not excited during the protocol execution,
the protocol attains energy transportation without heat generation in the channel.
We discuss the protocol focusing around qubit chains. In addition, we address a related problem of
breaking ground-state entanglement by measurements.
\end{abstract}
\end{titlepage}

\bigskip

\section{Introduction}

\ \newline

Recently protocols called quantum energy teleportation (QET) have been
proposed which transport energy by local operations and classical
communication (LOCC), respecting causality and local energy conservation. The
protocols can be considered for various many-body quantum systems, including
qubit chains \cite{hotta1,hotta2}, 1+1 dimensional massless Klein-Gordon
fields \cite{hotta3}, 1+3 dimensional electromagnetic field \cite{hotta5}, and
cold trapped ions \cite{hotta6}. The key point of the protocol is that there
exists quantum correlation between local fluctuations of different sites in
the ground state. The root of this correlation is the ground-state
entanglement. By virtue of the correlation, a measurement result of local
fluctuation in some site includes information about fluctuation in other
sites. By selecting and performing a proper local operation based on the
announced information, zero-point oscillation of a site far from the
measurement site can be more suppressed than that of the ground state,
yielding negative energy density. Here the origin of energy density is fixed
such that the expectational value vanishes for the ground state. Such negative
energy density appears due to quantum interference effects\ \cite{Ford}. Even
if we have a region with negative energy density in a system, we have other
regions with positive energy density and the total energy of the system
remains nonnegative. During the above local operation generating negative
energy density in the system, surplus energy is transferred from the quantum
fluctuation to external systems and can be harnessed.

The organization of this report is as followed: In section 2, the relation
between ground-state entanglement and emergence of negative energy density is
explained. QET is realized by generating negative energy density at a distant
site by LOCC. In section 3, a protocol of this QET is discussed for critical
Ising spin chains. In section 4, a related problem of breaking ground-state
entanglement by measurements are addressed. In section 5, recent results of
QET analysis are summarized for other quantum systems.

\section{Ground-State Entanglement and Negative Energy Density}

~

The QET protocol is able to work by virtue of ground-state entanglement and
emergence of negative energy density. In what follows, let us concentrate on
qubit chain systems and explain the entanglement and the negative energy
density. First of all, the Hamiltonian $H$ is given by a site sum of energy
density operators $T_{n}$, where $n$ denotes site number. The origin of
$T_{n}$ can be shifted so as to satisfy
\begin{equation}
\langle g|T_{n}|g\rangle=0\label{1}%
\end{equation}
without loss of generality. If each $T_{n}$ is a local operator at site $n$
satisfying $\left[  T_{n},~T_{n^{\prime}}\right]  =0$, all $T_{n}$ can be
simultaneously diagonalized. The ground state $|g\rangle$ becomes separable
and an eigenstate for the lowest eigenvalue of each $T_{n}$. Clearly, in such
a situation, $T_{n}$ is nonnegative. However, the condition $\left[
T_{n},~T_{n^{\prime}}\right]  =0$ is not sustained for cases with interactions
between qubits, and entangled ground states are generated. It is noted that a
correlation function $\langle g|T_{n}O_{m}|g\rangle$ of a separable ground
state $|g\rangle$ is given by $\langle g|T_{n}|g\rangle\langle g|O_{m}%
|g\rangle$ for a local operator $O_{m}$ at site $m$ apart far from $n$. On the
other hand, in the case of the entangled ground state $|g\rangle$, this
factorization relation does not hold in general:
\begin{equation}
\langle g|T_{n}O_{m}|g\rangle\neq\langle g|T_{n}|g\rangle\langle
g|O_{m}|g\rangle.\label{2}%
\end{equation}
This ground-state entanglement induces emergence of quantum states with
negative energy density as follows. It turns out first that the entangled
ground state $|g\rangle$ cannot be an eigenstate of $T_{n}$. The reason is
following. If the eigenvalue equation $T_{n}|g\rangle=\tau|g\rangle$ with a
real eigenvalue $\tau$ is satisfied, the above correlation function must be
written as
\[
\langle g|T_{n}O_{m}|g\rangle=\tau\langle g|O_{m}|g\rangle=\langle
g|T_{n}|g\rangle\langle g|O_{m}|g\rangle,
\]
where we have used $\langle g|T_{n}=\tau\langle g|$ and $\tau=\langle
g|T_{n}|g\rangle$. This obviously contradicts Eq. (\ref{2}). Therefore the
entangled ground state $|g\rangle$ satisfying Eq. (\ref{2}) is not an
eigenstate of $T_{n}$. Next let us spectral-decompose the operator $T_{n}$ as
\[
T_{n}=\sum_{\nu,k_{\nu}}\epsilon_{\nu}(n)|\epsilon_{\nu}(n),k_{\nu}%
(n)\rangle\langle\epsilon_{\nu}(n),k_{\nu}(n)|,
\]
where $\epsilon_{\nu}(n)$ are eigenvalues of $T_{n}$, $|\epsilon_{\nu
}(n),k_{\nu}(n)\rangle$ are corresponding eigenstates, and the index $k_{\nu
}(n)$ denotes the degeneracy freedom of the eigenvalue $\epsilon_{\nu}(n)$.
Because $\left\{  |\epsilon_{\nu}(n),k_{\nu}(n)\rangle\right\}  $ is a
complete set of \ orthonormal basis vectors of the total Hilbert space of the
qubit chain, the ground state can be uniquely expanded as
\[
|g\rangle=\sum_{\nu,k_{\nu}(n)}g_{\nu,k_{\nu}(n)}|\epsilon_{\nu}(n),k_{\nu
}(n)\rangle,
\]
where $g_{\nu,k_{\nu}(n)}$ are complex coefficients of the expansion. By use
of this expansion, Eq. (\ref{1}) gives an equation as follows:
\[
\langle g|T_{n}|g\rangle=\sum_{\nu,k_{\nu}(n)}\epsilon_{\nu}(n)\left\vert
g_{\nu,k_{\nu}(n)}\right\vert ^{2}=0.
\]
Clearly, this equation for $g_{\nu,k_{\nu}(n)}$ has no solution when the
lowest eigenvalue $\epsilon_{\min}(n)$ of $T_{n}$ is positive. The case with
$\epsilon_{\min}(n)=0$ is also prohibited for the equation because, if so, the
entangled ground state $|g\rangle$ would become an eigenstate of $T_{n}$ with
its eigenvalue $\tau=0$ and contradicts Eq. (\ref{2}), as proven above. This
means that $\epsilon_{\min}(n)$ must be negative. It is thereby verified that
there exist quantum states $|\epsilon_{\min}(n),k_{\min}(n)\rangle$ with
negative energy density due to the ground-state entanglement. Here it should
be stressed that, because of Eq. (\ref{1}), the eigenvalue of the ground state
is zero:
\[
H|g\rangle=0,
\]
and $H$ is a nonnegative operator. Therefore, even if we have a region with
negative energy density in a system, we have other regions with positive
energy density so as to make the total energy of the system nonnegative. In
the QET protocol, the negative energy density plays a crucial role as seen in
the next section.

\section{QET Protocol}

~

By use of the negative energy density, protocols of QET can be constructed. In
this section, a QET protocol for a critical Ising spin chain \cite{hotta2} is
explained. The Hamiltonian is given by a sum of energy density operator
$T_{n}$: $H=\sum_{n}T_{n}$. The operator $T_{n}$ is given by
\begin{equation}
T_{n}=-J\sigma_{n}^{z}-\frac{J}{2}\sigma_{n}^{x}\left(  \sigma_{n+1}%
^{x}+\sigma_{n-1}^{x}\right)  -\epsilon, \label{ising}%
\end{equation}
where $\sigma_{n}^{z}$ and $\sigma_{n}^{x}$ are Pauli matrices at site $n$,
$J$ and $\epsilon$ are real constants. By fine-tuning $\epsilon$, Eq.
(\ref{1}) is attained. The QET protocol is composed of \ the following three
steps: \ (i) For the ground state $|g\rangle$, an energy sender $A$ performs a
local measurement of $\sigma_{A}$ which is a one-direction component of the
Pauli spin operator acting on $A$'s qubit. Those eigenvalues of $\sigma_{A}$
are $(-1)^{\mu}$ with $\mu=0,1$. Let us write the spectral decomposition of
$\sigma_{A}$ as
\[
\sigma_{A}=\sum_{\mu=0,1}\left(  -1\right)  ^{\mu}P_{A}\left(  \mu\right)  ,
\]
where the operator $P_{A}\left(  \mu\right)  $ are projective operators onto
the eigenspaces. In this measurement process, $A$ must input positive amount
of energy given by
\[
E_{A}=\sum_{\mu=0,1}\langle g|P_{A}\left(  \mu\right)  HP_{A}\left(
\mu\right)  |g\rangle
\]
to the qubit chain. (ii) $A$ announces the measurement result $\mu$ to an
energy receiver $B$ by a classical channel. (iii) $B$ performs a local unitary
operation depending on the value of $\mu$. The unitary operator is defined by
\[
V_{B}\left(  \mu\right)  =I\cos\theta+i\left(  -1\right)  ^{\mu}\sigma_{B}%
\sin\theta,
\]
where $\sigma_{B}$ is a one-direction component of the Pauli spin operator
acting on $B$'s qubit, and the above real parameter $\theta$ is fixed so as to
extract the maximum energy from the chain. In this analysis, we assume that
dynamical evolution of the system induced by $H$ is negligible during short
time interval $t$ of the protocol: $\exp\left[  -itH\right]  $ $\sim I$.
\ Hence, the quantum state after step (iii) is written as follows.
\[
\rho=\sum_{\mu=0,1}V_{B}\left(  \mu\right)  P_{A}\left(  \mu\right)
|g\rangle\langle g|P_{A}\left(  \mu\right)  V_{B}^{\dag}\left(  \mu\right)  .
\]
Using this state, it can be shown that $B$ extracts positive energy $+E_{B}$
on average from the qubit chain, accompanied by excitations with negative
energy $-E_{B}$ in the qubit chain around $B$'s site in step (iii). In fact,
the expectational value of energy after step(iii) is calculated \cite{hotta1}
as%
\begin{equation}
\operatorname*{Tr}\left[  \rho H\right]  =E_{A}+\frac{\eta}{2}\sin\left(
2\theta\right)  +\frac{\xi}{2}\left(  1-\cos\left(  2\theta\right)  \right)  ,
\label{9}%
\end{equation}
where $\xi$ and $\eta$ are given by
\begin{align*}
\xi &  =\langle g|\sigma_{B}H\sigma_{B}|g\rangle\geq0,\\
\eta &  =i\langle g|\sigma_{A}\left[  H,~\sigma_{B}\right]  |g\rangle.
\end{align*}
The coefficient $\eta$ is a two-point correlation function of (semi-)local
operators of $A$ and $B$, and turns out to be real. It is a key point that
$\eta$ does not vanish in general because of the ground-state entanglement. By
taking a value of $\theta$ defined by
\[
\cos\left(  2\theta\right)  =\frac{\xi}{\sqrt{\xi^{2}+\eta^{2}}},~\sin
(2\theta)=-\frac{\eta}{\sqrt{\xi^{2}+\eta^{2}}},
\]
the minimum value of $\operatorname*{Tr}\left[  \rho H\right]  $ with respect
to $\theta$ is written explicitly as%

\[
\operatorname*{Tr}\left[  \rho H\right]  =E_{A}-\frac{1}{2}\left[  \sqrt
{\xi^{2}+\eta^{2}}-\xi\right]  .
\]
From the viewpoint of local energy conservation, this result implies that,
during the operation $V_{B}\left(  \mu\right)  $, positive amount of energy
given by
\begin{equation}
E_{B}=E_{A}-\operatorname*{Tr}\left[  \rho H\right]  =\frac{1}{2}\left[
\sqrt{\xi^{2}+\eta^{2}}-\xi\right]  >0 \label{eb}%
\end{equation}
is transferred from the qubit chain to external systems including the device
system executing $V_{B}\left(  \mu\right)  $. In addition, it is possible to
calculate analytically the value of $E_{B}$ for the critical Ising spin chain
as follows \cite{hotta2}.%

\begin{equation}
E_{B}=\frac{2J}{\pi}\left[  \sqrt{1+\left(  \frac{\pi}{2}\Delta(\left\vert
n_{A}-n_{B}\right\vert )\right)  ^{2}}-1\right]  , \label{101}%
\end{equation}
where $\Delta(n)$ is defined by
\[
\Delta(n)=\left(  \frac{2}{\pi}\right)  ^{n}\frac{2^{2n(n-1)}h(n)^{4}}{\left(
4n^{2}-1\right)  h(2n)}%
\]
with $h(n)=\prod_{k=1}^{n-1}k^{n-k}$. The asymptotic behavior of $\Delta(n)$
for large $n$ is given by
\begin{equation}
\Delta(n\sim\infty)\sim\frac{1}{4}e^{1/4}2^{1/12}c^{-3}n^{-9/4}, \label{20}%
\end{equation}
where the constant $c$ is evaluated as $c\sim1.28$. Due to the criticality of
this model, $E_{B}$ decays following not an exponential law but a power law
($\propto\left\vert n_{A}-n_{B}\right\vert ^{-9/2}$) for large separation.

\section{Breaking Ground-State Entanglement by Measurements}

~

In section 3, we have shown that $B$ obtains energy from the qubit chain by
the QET protocol. However, even after the last step (iii) of the protocol,
there exists residual energy $E_{A}$ that $A$ had to first deposit to the
qubit chain. Let us imagine that $A$ attempts to completely withdraw $E_{A}%
~$by local operations after step (iii). If $A$ succeeded in this withdrawing,
the energy gain of $B$ might have no cost. However, if so, the total energy of
the qubit chain became equal to $-E_{B}$ and negative. Meanwhile, we know that
the total energy of the qubit chain system must be nonnegative. Hence, $A$
cannot withdraw energy larger than $E_{A}-E_{B}$ by local operations at site
$n_{A}$. This means that, in the QET protocol, $B$ has borrowed energy $E_{B}$
in advance from the qubit chain on security of the deposited energy $E_{A}$.
The main reason for $A$'s inability to withdraw is because $A$'s local
measurement breaks the ground-state entanglement between $A$'s qubit and all
the other qubits. The post-measurement state is an exact separable state with
no entanglement. If $A$ wants to recover the original state of her qubit with
zero energy density, $A$ must recreate the broken entanglement. However,
entanglement generation needs nonlocal operations in general. Therefore, $A$
cannot recover the state perfectly by her local operations alone. This
interesting aspect poses a residual-energy problem of the ground-state
entanglement broken by measurements. Let us imagine that $A$ stops the QET
protocol soon after step (i) of the protocol, and attempts to completely
withdraw $E_{A}~$by local operations. By the same argument as the above, \ it
is shown that this attempt never succeeds because $A$ breaks the ground-state
entanglement. Of course, for a long time interval beyond the short time scale
that we have considered, local cooling is naturally expected to make residual
energy in the qubit chain approaching zero by an assist of dynamical evolution
induced by the nonlocal Hamiltonian $H$. However, in this short time interval,
the dynamical evolution is not available. Therefore it is concluded that the
residual energy in the qubit chain has its nonvanishing minimum value $E_{r}$
with respect to $A$'s local cooling processes in short time. In order to make
the argument more concrete, let us consider a general local cooling operation
of $A$ after step (i) obtaining the measurement result $\mu$. The operation is
expressed by use of $\mu$-dependent Kraus operators $M_{A}(\alpha,\mu)$
satisfying%
\begin{equation}
\sum_{\alpha}M_{A}^{\dag}(\alpha,\mu)M_{A}(\alpha,\mu)=I. \label{100}%
\end{equation}
Then the quantum state after this local cooling by $A$ is given by%
\begin{equation}
\rho_{c}=\sum_{\mu,\alpha}M_{A}(\alpha,\mu)P_{A}\left(  \mu\right)
|g\rangle\langle g|P_{A}\left(  \mu\right)  M_{A}^{\dag}(\alpha,\mu).
\label{31}%
\end{equation}
The minimum value $E_{r}$ of the residual energy with respect to $M_{A}%
(\alpha,\mu)$ satisfying Eq. (\ref{100}) is written as%

\begin{equation}
E_{r}=\min_{\left\{  M_{A}(\alpha,\mu)\right\}  }\operatorname*{Tr}\left[
\rho_{c}H\right]  . \label{40}%
\end{equation}
Evaluation of $E_{r}$ is performed analytically in the Ising spin chains
\ \cite{hotta2} and given by%

\[
E_{r}=\left(  \frac{6}{\pi}-1\right)  J>0,
\]
for the critical chain.~Surprisingly, $A$ is not able to extract this energy
by any local operation in the short time, though it exists in front of $A$.
Because of the nonnegativity of $H$, it is easily checked by resuming the QET
protocol after the local cooling that $E_{r}$ is lower bounded by the
teleported energy $E_{B}$ in Eq. (\ref{eb}). In addition, the paper
\cite{hotta2} gives a stringent argument that the teleported energy in an
extended protocol gives a more tight lower bound of residual energy $E_{r}$
for general qubit chains.

Finally, a comment is added about recent numerical researches of the
ground-state entanglement. As a quantitative entanglement measure, the
negativity has been computed between separated blocks of qubit chains
\cite{bose} ( the logarithmic negativity for harmonic oscillator chains
\cite{reznik}) showing that at criticality this negativity is a function of
the ratio of the separation to the length of the blocks and can be written as
a product of a power law and an exponential decay. In our setting of QET, this
suggests that change of the entanglement between $A$'s block and $B$'s block
after $A$'s local measurement has a similar rapid-decay dependence on the
separation with a fixed block length. Thus it may be concluded that the
entanglement between $A$'s block and $B$'s block itself is not essential for
QET. Though the entanglement between the two blocks may be rapidly damped,
$E_{B}$ shows a power law decay ($\propto n^{-9/2}$) for large separation $n$,
as seen in Eq. (\ref{101}) and Eq. (\ref{20}). This implies in a sense that
almost "classical" correlation between $A$'s block and $B$' block is
sufficient to execute QET for large separation, and is expected to be robust
against environment disturbance, contrasting to the entanglement fragility. It
should be emphasized, however, that this "classical" correlation is originally
induced by the ground-state entanglement characterized by Eq. (\ref{2}). If
the ground state is separable, we have no correlation between the blocks.

\bigskip

\section{QET for Other Systems}

~

The QET protocols can be considered for other quantum systems. In
\cite{hotta3}, a protocol of QET for 1+1 dimensional massless scalar fields is
analyzed. Though the nonrelativistic treatment for the qubit chain in section
3 is valid for short-time-scale processes of QET in which dynamical evolution
induced by the Hamiltonian is negligible, in this relativistic case, the
dynamical effect propagates with light velocity, which is the upper bound on
the speed of classical communication. Thus, we generally cannot omit global
time evolution. It is also noted that any continuous limit of zero lattice
spacing cannot be taken for the protocols in the lattice QET models as long as
measurements in the protocols are projective, which becomes an obstacle to
obtaining a smooth limit. Therefore, in \cite{hotta3}, $A$ makes not a
projective but instead a well-defined POVM measurement to the vacuum state of
the field. After wavepackets with light velocity excited by $A$'s measurement
have already passed by the position of $B$, $B$ extracts energy from the local
vacuum state of the field by a unitary operation dependent on the measurement
result announced by $A$. In \cite{hotta5}, two QET protocols with discrete and
continuous variables are analyzed for 1+3 dimensional electromagnetic field.
In the discrete case, a 1/2 spin is coupled with the vacuum fluctuation of the
field and measured in order to get one-bit information about the fluctuation.
In the continuous case, a harmonic oscillator is coupled with the fluctuation
and measured in order to get continuous-variable information about the
fluctuation. In the discrete case, the amount of the extracted energy is
suppressed by an exponential damping factor when the energy infused by the
measurement becomes large. This suppression factor becomes power damping in
the continuous case, and it is concluded that more information about the
vacuum fluctuation is obtained by the measurement, more energy can be
teleported. In \cite{hotta6}, a protocol of QET is proposed for trapped ions.
$N$ cold$~$ions, which are strongly bound in the $y$ and $z$ directions but
weakly bound in an harmonic potential in the $x$ direction, form a linear ion
crystal. The first ion that stays at the left edge of the crystal is the
gateway of the QET channel where energy is input. The $N$-th ion that stays at
the right edge of the crystal is the exit of the QET channel where the
teleported energy is output. Two internal energy levels of the gateway ion are
selected and regarded as energy levels of a probe qubit to measure the local
phonon fluctuation. The probe qubit is strongly coupled with the phonon
fluctuation in the ground state during short time via laser field and is
projectively measured. In the measurement models, the kinetic energy of the
gateway ion increases after the measurement, but the kinetic energy of other
ions and the potential energy of all the ions remain unchanged. The obtained
information is announced through a classical channel from the gateway point to
the exit point. The speed of the information transfer can be equal to the
speed of light in principle, which is much faster than that of the phonon
propagation\ in the ion crystal. The phonons excited at the QET gateway do not
arrive at the exit point yet when the information arrives at the exit point.
However, by using the announced information, we are able to soon extract
energy from the exit ion. Experimental verification of the QET mechanism has
not been achieved yet for any system, and is a quite stimulating open problem.

\bigskip

\textbf{Acknowledgments}\newline

\bigskip

I would like to M. Ozawa and A. Furusawa for fruitful discussions. This
research is partially supported by the SCOPE project of the MIC and the
Ministry of Education, Science, Sports and Culture of Japan, No. 21244007.

\end{document}